\documentstyle[prl,preprint,aps]{revtex}
\draft
\begin{document}

\title{Elastic theory of pinned flux lattices}
\author{Thierry Giamarchi}
\address{Laboratoire de Physique des Solides, Universit{\'e} Paris--Sud,
                   B{\^a}t. 510, 91405 Orsay, France\cite{junk} \\
                   email: giam@solrt.ups.circe.fr}
\author{Pierre Le Doussal\cite{frad}}
\address{Serin Physics Laboratory,
         Rutgers University, Piscataway NJ 08855, U.S.A. \\
         email: ledou@physics.rutgers.edu}
\maketitle

\begin{abstract}
The pinning of flux lattices by weak impurity disorder is studied
in the absence of free dislocations using
both the gaussian variational method and, to $O(\epsilon=4-d)$,
the functional renormalization group.
We find universal logarithmic growth of displacements for $2<d<4$:
$\overline{\langle u(x)-u(0) \rangle ^2}\sim A_d \log|x|$
and persistence of algebraic quasi-long range translational order.
When the two methods can be compared they agree within $10\%$ on
the value of $A_d$.
We compute the function describing the crossover between the
``random manifold'' regime
and the logarithmic regime. This crossover should be observable
 in present decoration experiments.
\end{abstract}
\narrowtext

It has been argued on the basis of various elastic models for
vortex lattices,
such as Larkin's model of independent
random forces acting on each vortex, that arbitrarily weak disorder
destroys translational order below four dimensions\cite{larkin_70}.
There is considerable disagreement, however, on the exact behaviour
of the density-density correlation function $C_K(r)=\langle\rho_K(0)
\rho_K(r)\rangle$
even in the
simpler case where dislocations are excluded. While direct
extensions of Larkin's model predict exponential decay in $d=3$
\cite{chudnovsky_pinning},
it has been pointed out that beyond the Larkin-Ovchinikov pinning length
$L_c$ the lattice behaves collectively as an elastic manifold in
a random potential with many metastable states, leading to a
different power-law roughening of the lattice and
to stretched exponential decay of $C_K(r)$\cite{feigelman_collective,%
bouchaud_variational_vortex_prl,bouchaud_variational_vortex}. On the
other hand, Flory-type arguments were
proposed making explicit use of the periodicity of the lattice
leading to logarithmic roughening \cite{nattermann_pinning}.
The role of dislocations
at weak disorder
above two dimensions is presently unsettled, but
Bitter decoration experiments \cite{grier_ecoration_manips} show
remarkably large regions free of dislocations and provide a strong motivation
for a better understanding of the elastic model.
Other related pinned elastic systems such as charge density waves,
magnetic bubbles, Wigner crystal, are under current active experimental
 study \cite{gruner_revue_cdw,shesadri_bubbles}.

In this Letter we take into account both the existence of many metastable
 states and the periodicity of the lattice.
We are primarily interested in the triangular Abrikosov lattice ($d=2+1$).
We also mention the case of $d=2+0$ (point vortices in thin films)
or $d=1+1$ (lines in a plane).  We show that
in the absence of dislocations, the translational correlation function
 has a slow algebraic decay in dimension larger than two, and thus
quasi-long range order persists. Two important length scales control
 the crossover towards this
asymptotic decay.
i) When the mean square of the relative displacement
 $\tilde{B}(x)=\frac12{\overline{\langle [u(x)-u(0)]^2 \rangle}}$ of
two lines as a function of their separation $x$ is shorter than the
 square of the Lindemann length $l_T^2 = \langle u^2 \rangle$, the
 thermal wandering of the
lines averages enough over the random potential and the model becomes
 equivalent to the random force
Larkin model for which $\tilde{B}(x) \sim |x|^{4-d}$. At low enough
 temperature, $l_T$ is replaced by the superconducting coherence length
 $\xi_0$ (i.e. the
correlation length of the random potential \cite{feigelman_collective,%
bouchaud_variational_vortex_prl,bouchaud_variational_vortex}).
ii) For $l_T^2 \ll \tilde{B}(x) \leq a^2$,
$\tilde{B}(x) \sim x^{2\nu}$: this is the random manifold regime where
each line sees effectively an independent random potential.
iii) For $x>\xi$, where $\xi$ corresponds to a relative displacement
 of the order of the lattice spacing $a$, $\tilde{B}(x=\xi) \sim a^2$,
the periodicity of the lattice becomes important.
We find $\tilde{B}(x) \sim A_d \log|x|$ where $A_d$ is a universal
 amplitude depending on dimension only and isotropy is recovered at
large distances. This leads to quasi long range order $C_{K_0}(r) \sim (1/r)^{
A_d}$.
We have computed the full crossover function in $d=3$ (Fig. 1).
It suggests that all the above regimes could be observed by analysis
 of the dislocation-free decoration samples.
These results are obtained using the Mezard-Parisi variational
 method \cite{mezard_variational_replica} first applied in this
 context by Bouchaud, Mezard and Yedidia (BMY)
\cite{bouchaud_variational_vortex_prl,bouchaud_variational_vortex}.
 Our results are at variance with BMY, for reasons detailed below.
In addition, we perform an $\epsilon = 4-d$ expansion using the
 functional renormalization group. The amplitudes $A_d$ obtained
 by these two rather different methods agree at order $\epsilon$ within 10\%.
In $d=2$, thermal fluctuations are important ($l_T=\infty$) and the
random manifold regime is much reduced.
We find a modified Larkin regime with T-dependent exponents
$\tilde{B}(x) \sim |x|^{2\nu(T)}$ and a long
distance logarithmic regime.  Details can be found in
\cite{ledoussal_variational_long}.

We denote
by $R_i$ the equilibrium position of the lines labeled by an integer $i$,
 in the $x-y$ plane, and by $u(R_i,z)$ their in-plane displacements.
 $z$ denotes the coordinate perpendicular to the planes.
 For weak disorder $a/\xi \ll 1$
it is legitimate to assume that $u(R_i,z)$ is slowly
varying on the scale of the lattice and to use a continuum elastic energy,
 as a function of the continuous variable $u(x,z)$. Impurity disorder is
modeled by
a gaussian random potential $V(x,z)$ with correlations:
$\overline{ V(x,z)V(x',z') }=\Delta(x-x')\delta(z-z')$ where $\Delta(x)$
is a short range function of range $\xi_0$ and Fourier transform $\Delta_q$.
The total energy is:
\widetext
\begin{equation} \label{total}
H_{\rm el}= \frac12 \int d^2x dz [(c_{11}-c_{66})(\partial_\alpha u_\alpha)^2
+c_{66} (\partial_\alpha u_\beta)^2 + c_{44}
(\partial_z u_\alpha)^2]  + \int d^2x dz V(x,z) \rho(x,z)
\end{equation}
\narrowtext
where $\alpha,\beta$ denote in-plane coordinates and
the density is $\rho(x,z) = \sum_i \delta(x - R_i -u(R_i,z))$.
Although we have also performed the calculations directly on the
Hamiltonian (\ref{total}) \cite{ledoussal_variational_long} it is
 more enlightening to use the following decomposition of the density
 that keeps track of the discreteness of the
lines. In the absence of dislocations, generalizing \cite{haldane_bosons},
one
introduces
the slowly varying field $\phi(x,z) = x - u(\phi(x,z),z)$.
 The density can be rewritten as
$\rho(x,z) =  \rho_0 {\rm det}[\partial_\alpha \phi_\beta] \sum_K
e^{i{K}\cdot\phi(x,z)}
\simeq
\rho_0( 1 - \partial_\alpha u_\alpha (\phi(x,z),z) +
\sum_{K \ne 0} e^{i {K} \cdot {x}} \rho_K(x))$, where $\rho_K(x)=
e^{-i K \cdot {u}(\phi(x,z),z)}$ is the usual translational order parameter
defined in terms of the reciprocal lattice vectors $K$, and
$\rho_0$ is the average density.

Using the replica trick on (\ref{total})
the disorder term gives $ - 1/(2T) \sum_{a,b} \int d^2x d^2 x' dz \Delta(x-x')
\rho^a(x,z) \rho^b(x',z)$.
 The above decomposition for the density leads to our starting model:
\widetext%
\begin{eqnarray} \label{cardyos}
H_{\text{eff}}& = & \int \frac{d^2q dq_z}{2 (2\pi)^3}  \sum_{a}
G_{0,\alpha\beta}^{-1}
u_\alpha^a(q,q_z) u_\beta^a(q,q_z) \\
 & & - \int d^2x dz \sum_{a,b} [\frac{\rho_0^2\Delta_0}{2T}\partial_\alpha
 u_\alpha^a\partial_\beta u_\beta^b
+ \sum_{K\ne 0} \frac{\rho_0^2\Delta_K}{2T}
\cos(K \cdot (u^{a}(x,z)-u^{b}(x,z)))] \nonumber
\end{eqnarray}
\narrowtext%
with $G_0^{-1} = (c_{44} q_z^2
+ c_{66} q^2)P^T_{\alpha\beta} + (c_{44} q_z^2
+ c_{11} q^2)P^L_{\alpha\beta}$ in the case of (\ref{total}), where
 $P^T_{\alpha\beta} = \delta_{\alpha\beta} - q_\alpha q_\beta / q^2$
and $P^L_{\alpha\beta} = q_\alpha q_\beta / q^2$.
To be rigorous (\ref{cardyos}) should be written in term of $u(\phi(x,z),z)$.
This however has no effect on our results. It leads only to
corrections of higher order in $\nabla  u$ which we neglect since
we work here in the elastic limit $a/\xi \ll 1$
\cite{ledoussal_variational_long}.
For clarity we present the calculation for the
isotropic version of (\ref{cardyos}) in $d$ dimensions with
 $G_0^{-1} = c q^2 \delta_{\alpha\beta}$ where $c$ is an elastic
 constant and $\rho_0^2 \Delta_K = \Delta$ for all $K$ ($x \equiv (x,z)$).
The results for (\ref{total}) is presented at the end.
A related single cosine model was studied by Villain and Fernandez
 \cite{villain_cosine_realrg} using a real space RG. Our results
 confirm and extend their analysis.

One can get an idea of the effect
of various terms in (\ref{cardyos}), by using arguments similar
to \cite{fukuyama_pinning}. Assuming that
$u$ varies of $\sim a$ over a length $\xi$, the density of kinetic
energy is $\sim c(a/\xi)^2$ . The long
wavelength part of the disorder is
$H^{\text{dis}}_{q \sim 0} \sim \int d^dx V(x) \partial_\alpha
  u_\alpha(x)\sim \Delta^{1/2} /\xi^{1+d/2}$
and
$H^{\text{dis}}_{q \sim K_0}
                      \sim \int d^dx V(x) \cos(K_0 \cdot
                      (x- u(x)) \sim \Delta^{1/2} /\xi^{d/2}$
where $K_0$ is the first reciprocal lattice vector.
There are thus two length scales $\xi_{q \sim 0} \sim
 a\left(c^2 a^{d}/\Delta\right)^{1/(2-d)}$ and
$ \xi \sim a \left(c^2 a^{d}/\Delta\right)^{1/(4-d)}$.
The $q\sim 0$ component of the disorder is therefore relevant only for
$d \leq 2$ and the second term in (\ref{cardyos}) can be dropped for $d>2$.
Higher Fourier components $V_{q \sim K}$ disorder the lattice below $d=4$.

We now look for the best trial Gaussian Hamiltonian $H_0$ in
replica space, of the form \cite{mezard_variational_replica}:
\begin{equation} \label{variat}
H_0 = {1 \over 2} \int {d^dq \over (2 \pi)^d} G^{-1}_{ab}(q)
u_a(q) \cdot u_b(-q)
\end{equation}
Defining the self-energy $G^{-1}_{ab}=  c q^2 \delta_{ab} -
\sigma_{ab}$, and $G_c^{-1}(q) = \sum_b G_{ab}^{-1}(q)$ we obtain by
minimization
of the variational free energy $F_{\text{var}}=F_0+\langle H-H_0
\rangle_{H_0}$ the saddle point equations:
\begin{eqnarray} \label{bebe}
G_c(q) & = & \frac1{c q^2}, \quad \sigma_{a \ne b} =  {\Delta \over n T}
 \sum_{K} K^2 e^{- {K^2 \over 2}  B_{ab}(x=0)} \\
B_{ab}(x) & = & \frac1n \langle [u_a(x) - u_b(0)]^2 \rangle  \\
            & = &  T \int {d^dq \over {(2\pi)}^d}
(G_{aa}(q) + G_{bb}(q) - 2 \cos(qx) G_{ab}(q)) \nonumber
\end{eqnarray}
where $n$ is the number of components of $u$.

The replica symmetric solution $B_{a \ne b}(x)=B(x)$, which mimics the
 distribution of displacements by a single
Gaussian, is always unstable for $2<d<4$. This is seen from
the eigenvalue $\lambda$ of the replicon mode
\cite{mezard_variational_replica}:
\begin{equation}
\lambda=1 - \frac{\Delta}{n} \sum_K
K^4 e^{- K^2 T \int {d^dp \over (2\pi)^d}
G_c(p) } \int {d^dq \over (2\pi)^d} G_c^2(q)
\end{equation}

Introducing a small regularizing mass in $G_c$: $G_c(q)^{-1}=c q^2 + \mu^2$
we find, when $\mu \to 0$, that for $d<2$ the replica symmetric solution
is always stable and disorder is irrelevant.
For $d=2$ the condition becomes $\mu^{-2(1-{T K_0^2 \over 4 \pi c})}<1$ for
small $\mu$.
Thus there is a transition
at $T=T_c=4 \pi c/K_0^2 $ between a stable high-T phase where disorder is
irrelevant and a low-T glassy phase
where the symmetric saddle point is unstable. For $2<d<4$ it is {\it always
unstable} and disorder is always relevant.

We now find a replica symmetry breaking solution for
$2<d<4$, the $d=2$ case being discussed later. Following
\cite{mezard_variational_replica} we denote $\tilde{G}(q)=G_{aa}(q)$
and parametrize $G_{ab}(q)$
by $G(q,v)$ where $0<v<1$, and similarly for $B_{ab}(x)$.
The
saddle point equations become:
\begin{equation} \label{saddle}
\sigma(v)= \frac{\Delta}{n T} \sum_K  K^2 e^{-{1\over 2} K^2 B(0,v)}
\end{equation}
Expression (\ref{bebe}) and
the algebraic rules for inversion of hierarchical matrices
\cite{mezard_variational_replica} give:
\begin{equation} \label{inversion}
B(0,v)=B(0,v_c)+ \int_{v}^{v_c} dw \int \frac{d^dq}{(2 \pi)^d}
{ 2 T \sigma'(w) \over {(G_c(q)^{-1} + [\sigma](w))}^2 }
\end{equation}
where $[\sigma](v)=u\sigma(v)-\int_{0}^{v} dw \sigma(w)$ and
$v_c$ is the breakpoint such that $\sigma(v)$ is constant for $v>v_c$.
$B(0,v_c)$ is a nonuniversal quantity $B(0,v_c) \simeq \xi_0^2 + l_T^2$.

To discuss the large distance behaviour $x \gg \xi$ it is enough
 to keep $K^2=K_0^2$ in (\ref{saddle}) since $B(0,v) \gg a^2$.
 In that case,
taking the derivative of (\ref{saddle}) with respect to $v$, using
$[\sigma]'(v)=v\sigma'(v)$ and (\ref{inversion}) one finds
the effective self energy:
\begin{equation}\label{sigmeq}
[\sigma](v)=  (v/v_0)^{2 / \theta}
\end{equation}
and $v_0 = 2 K_0^2 T c_d c^{-d/2}/(4-d)$. The energy fluctuation
exponent is $\theta=d-2$. Energy fluctuations are of order $T/v$ and
the large scale behaviour is controlled
by small $v$. One can now compute the correlation functions:
\begin{eqnarray} & &
\overline{ \langle (u(x) - u(0))^2 \rangle } = 2 n T \int {d^dq \over (2\pi)^d}
(1 - \cos(qx) ) \tilde{G}(q) \\ & &
\tilde{G}(q) = { 1\over{c q^2}} ( 1 + \int_{0}^{1} {dv \over v^2}
{ [\sigma](v) \over { c q^2 + [\sigma](v) } } ) \sim \frac{Z_d}{q^d}
\end{eqnarray}
with $Z_d=(4-d)/(TK_0^2 S_d)$ and $1/S_d=2^{d-1}\pi^{d/2}\Gamma[d/2]$.
Thus for $2<d<4$ we find {\it logarithmic} growth:
\begin{equation} \label{meansq}
\overline{ \langle (u(x) - u(0))^2 \rangle }=\frac{2n}{K_0^2} A_d \log|x|
\end{equation}
with $A_d=4-d$.

We now give the full crossover function for $d=3$  for model (\ref{cardyos})
 with $\rho_0^2 \Delta_{K_0}= \Delta$. The crossover length is
 $\xi = a^4 c^2/(2 \pi^3 \Delta)$.
Defining $h(z) = \sum_P (\Delta_K/\Delta_{K_0}) P^4 e^{-z P^2}$,
where $K=2 \pi P/a$,
the solution is, in parametric form:
\begin{equation} \label{crosso}
v = v_\xi h'(z) \qquad [\sigma]^{1/2} = c^{1/2} \xi^{-1}  h(z)
\end{equation}
with $B(0,v) = a^2 z/(2\pi^2)$ and
$v_\xi=2\pi^4 T \Delta/(a^6 c^3)$ $\sim$ $l_T/\xi$. The mean square
displacement $\tilde B$ is:
\begin{eqnarray} \label{bani}
& \tilde{B}(x) = \frac{a^2}{2\pi^2} \tilde{b}( \frac{x}{\xi}) \quad
\tilde{b}(x)
= \int_0^\infty dt \frac{h''(t) h(t)}{h'(t)^2} f(x h(t))  &  \\
& f(x) = 1 - \frac1x (1 - e^{-x})  &
\end{eqnarray}
For $x \ll \xi$, $B(0,v)$ is very small and $h(z) \sim 1/z^{n/2+2}$
and therefore $\tilde{B}(x) \sim x^{2 \nu}$ with $\nu=1/6$ for $n=2$.
This corresponds to the random manifold regime
\cite{bouchaud_variational_vortex_prl,bouchaud_variational_vortex}.
 The crossover function (\ref{crosso}-\ref{bani}) was
 derived assuming $v_\xi \ll v_c$, equivalent to $l_T \ll \xi$.
At scales such that $\tilde{B}(x)$ is smaller than $l_T^2$ or $\xi_0^2$
one is in the regime $v > v_c$
and one recovers the replica symmetric propagator $\tilde{G}(q)
 \sim 1/q^4$ for $q^2 \gg [\sigma](v_c)$, and Larkin's model behavior.

In the vortex lattice (\ref{total}), shear deformations dominate
($c_{66} \ll c_{11}$) and
the crossover length becomes
$\xi = c_{66}^{3/2} c_{44}^{1/2} a^4 / (2 \pi^3 \Delta)$.
$\tilde{B}_{\alpha\beta}(x)=\tilde{B}_T P^T_{\alpha\beta}(x)+
\tilde{B}_L P^L_{\alpha\beta}(x)$ depends on the direction
of $x$ and we find:
\begin{equation} \label{exact}
\tilde{B}_{L,T}(x) =  \frac{a^2}{2 \pi^2}[ \tilde{b}_{T,L}(\frac{x}{\xi})
+ \frac{c_{66}}{c_{11}}
\tilde{b}_{L,T}((\frac{x}{\xi} \sqrt{\frac{c_{66}}{c_{11}}})]
\end{equation}
where $b_{L,T}$ have an expression similar to (\ref{bani}) with $f$ replaced by
\begin{equation}
f_{L}(x) = \frac12 - \frac1{x^2} + (\frac1x + \frac1{x^2}) e^{-x} \quad
f_{T}(x) = f(x) - f_L(x)
\end{equation}
The crossover function is shown
in Fig.~\ref{figure1}. We find that if $\xi/a \gg 1$ all curves
should scale when plotted in units of $x/\xi$.
Such a crossover should be observable in decoration experiments.
 The ratio $R=\tilde{B}_T(x)/\tilde{B}_L(x)$ crosses over at $x=\xi$ from
$R=4/3$ towards $R=1$ as isotropy is restored at large scale.

As $d \rightarrow 2^+$
the function $[\sigma](v)$ in (\ref{sigmeq}) vanishes
for $v<v_0=T K_0^2/(4 \pi c)$. Thus in $d=2$ for $T<T_c$ there is a
one-step replica symmetry broken solution with
$[\sigma](v)=0$ for $v<v_c$ and $[\sigma](v)=[\sigma](v_c)$
for $v_c<v<1$ \cite{ledoussal_variational_long}. In the glassy
phase $T<T_c$ the variational
method predicts the following two regimes. For $T$ moderate, the
 short distance Larkin regime $\tilde{B}(x) \sim x^2$ crosses over directly
towards the asymptotic logarithmic growth (\ref{meansq}) at a length $\xi =
a ( c^2 a^2/\Delta )^{1/(2-T K_0^2/2 \pi c)}$. At low $T$, $T
\log(\xi/a)/(ca^2) \ll 1$,
there is, in addition, an intermediate random manifold regime.
In $d=1+1$ the starting model (\ref{cardyos}) becomes exact due
to the absence of dislocations and a RG calculation
for $T \approx T_c$ \cite{villain_cosine_realrg,cardy_desordre_rg,toner_log_2}
 finds $\tilde{B}(x) \sim \log^2(x)$ at large $x$.
If one believes in the validity of this RG in a glassy phase, the
simple gaussian ansatz does not give the exact long range behaviour,
due to the importance of fluctuations in $d=2$. However it gives $T_c$ exactly
and captures
correctly the crossover towards a slower logarithmic regime. Using RG we have
shown
\cite{ledoussal_variational_long} that in $d=2$ the Larkin regime is
in fact {\it anomalous}
with a continuously variable exponent:
\begin{equation}
\tilde{B}(x) \sim a^2 ( x/\xi )^{2 - \frac{T K_0^2}{2 \pi c}}
\end{equation}
for $x<\xi$ ( in the low-T regime $\xi$ is replaced by another length).
Model (\ref{cardyos}) will probably apply in $d=2+0$ at scales shorter
 than the distance between dislocations. It could also describe a
polymerized membrane on a disordered substrate, with very high
dislocation core energy.

A previous application of the variational method by BMY
\cite{bouchaud_variational_vortex_prl,bouchaud_variational_vortex}
led to the conclusion which we believe is erroneous,
that the fluctuations are enhanced at large distances. They applied
 the same method to a model in which each line
sees a different disorder. This amounts to introduce an extra and unphysical
disorder in the original model (\ref{total}) with correlations decaying
as $1/|R_i-R_j|^\lambda$.
The long wavelength part of such a disorder dominates
since large global translations of the lattice can
improve the bulk energy, whereas in the physical model
the gain in energy from the long wavelengths components can only come
 from surface terms and is thus irrelevant for $d>2$, as shown in the
discussion
following (\ref{cardyos}). Indeed
for $\lambda \to 0$ the amplitude they obtain vanishes.
In fact one can simplify the
the saddle point equations of
\cite{bouchaud_variational_vortex_prl,bouchaud_variational_vortex}
by noting that the $x$ dependence of $B(x,u)$ in these equations is
unimportant,
 up to higher order terms in $\nabla u$. The resulting local
 model (\ref{cardyos}) is simple enough to allow for the exact solution
(\ref{exact}).

To complement the variational method we perform a functional renormalization
group
 calculation on the isotropic model.
To simplify we take $u$ to be a scalar field ($n=1$) and set $c=1$.
One defines the replicated Hamiltonian as:
\begin{equation}
H_{\rm imp} = - \frac1{2T} \sum_{ab} \int d^dx \Delta(u_a(x) - u_b(x))
\end{equation}
The function $\Delta(z)$ is periodic of period $1$ ($2 \pi K_0=1$).
The RG equations to order $\epsilon = 4-d$ have been derived by D.S. Fisher
 \cite{fisher_functional_rg} for the random manifold problem:
\begin{equation}
\frac{d \Delta}{dl} = (\epsilon - 4 \zeta) \Delta  + \zeta z \Delta'
+ \frac12 (\Delta'')^2 - \Delta'' \Delta''(0)
\end{equation}
A factor $S_d$ has been absorbed into $\Delta$.
The periodicity of the function implies that the roughening exponent
is $\zeta=0$. This allows to obtain the fixed point function
$\Delta^*(z)$ in the interval $[0,1]$:
\begin{equation}
\Delta^*(z) = \frac{\epsilon}{72} (\frac1{36} -z^2(1-z)^2)
\end{equation}
Values for other $z$ are obtained by periodicity. The fixed point is
 stable except for a constant shift. The linearized spectrum is discrete
 and the eigenvectors are Jacobi polynomials.
This function has a nonanaliticity \cite{fisher_functional_rg}
 $\Delta^{*(4)}(0) = \infty$.
To compute the correlation function $\tilde{\Gamma} = T \tilde{G}$
one uses the RG flow equation:
\begin{equation}
\tilde{\Gamma}(q,T,\Delta) = e^{dl} \tilde{\Gamma}(q e^l,T e^{(2-d)l},
\Delta(l))
\end{equation}
choosing $e^{l^*} q = 1/a$ allows to obtain perturbatively
 $\tilde{\Gamma}(q) = -\Delta''^*(0) / q^4 = a^{(4-d)} \epsilon / (36 S_d
q^d)$.
 Thus we find, at order $\epsilon$, a logarithmic growth (\ref{meansq})
 of line displacement
with $A_{d,RG} = \epsilon (2 \pi)^2/36=1.10\epsilon$. This compares
within 10\% with $A_{d,VAR} = \epsilon$ which slightly underestimate
 fluctuations. For $2 \le d < 4$ the real space RG method
of\cite{villain_cosine_realrg} also predicts a $\log$
 but does not allow to compute the universal prefactor $A_d$ or
 the crossover function. The agreement between these methods,
 none being rigorous, lends credibility to
the additional results in $d=3$ obtained using the variational method.

To conclude, we have shown that due to the periodicity of the lattice, the
pinning by impurities becomes less effective at large length scales. This
has consequences on the transport properties. In the flux creep regime
energy barrier arguments \cite{feigelman_collective}, and the exponents found
here,
 lead to a non linear voltage-current relation
$V \sim \exp[-1/(Tj^{\mu})]$ where $\mu$ crosses
over from $\mu \approx 0.7-0.8$ to $\mu=1/2$ as $j$ decreases.

We thank D.S. Fisher, T. Hwa and D. Huse for interesting
discussions. Both
authors thank AT\&T Bell Laboratories, where part of this work was completed,
for support and hospitality. PLD acknowledges support from NSF Grant 4-20508.

%\bibliographystyle{prsty}
%\bibliography{revues,global,creep}

\newpage
\begin{figure}
\caption{\label{figure1}
Plot of $\tilde{b}_L$ versus $\log|x/\xi|$ ( solid line ), as defined in the
text, for the Abrikosov triangular lattice.
The dashed line is the random manifold result.}
\end{figure}
\end{document}